\documentclass[aps,prl,twocolumn,amsmath,amssymb,floatfix,longbibliography,superscriptaddress]{revtex4-2}
\usepackage{physics}
\usepackage{mathrsfs}
\usepackage{graphicx}
\usepackage{dcolumn}
\usepackage{amssymb}
\usepackage{bm}
\usepackage{empheq}
\usepackage{pifont}
\usepackage[hidelinks]{hyperref}
\usepackage{xcolor}
\hypersetup{
    colorlinks,
    linkcolor={blue},
    citecolor={blue!80!black},
    urlcolor={blue!80!black}
}
\usepackage[toc,page]{appendix}
\usepackage[normalem]{ulem}
\usepackage{epstopdf}
\usepackage[separate-uncertainty]{siunitx}
\usepackage{upgreek}
\begin{document}

\title{Magnetic island structures in relativistic laser-driven plasma channels}
\author{Dongchi Cai}
\affiliation{State Key Laboratory of Nuclear Physics and Technology, and Key Laboratory of HEDP of\\ the Ministry of Education, CAPT, School of Physics, Peking University, Beijing 100871, China}
\affiliation{Institute of Theoretical Physics, Chinese Academy of Sciences, Beijing 100190, China}
\author{Zheng Gong}
\email{zgong92@itp.ac.cn}
\affiliation{Institute of Theoretical Physics, Chinese Academy of Sciences, Beijing 100190, China}
\author{Guanqi Qiu}
\affiliation{State Key Laboratory of Nuclear Physics and Technology, and Key Laboratory of HEDP of\\ the Ministry of Education, CAPT, School of Physics, Peking University, Beijing 100871, China}
\author{Deji Liu}
\affiliation{State Key Laboratory of Nuclear Physics and Technology, and Key Laboratory of HEDP of\\ the Ministry of Education, CAPT, School of Physics, Peking University, Beijing 100871, China}
\author{Yinren Shou}
\affiliation{Institute of Modern Physics, Fudan University, Shanghai 200433, China}
\author{Xueqing Yan}
\email{x.yan@pku.edu.cn}
\affiliation{State Key Laboratory of Nuclear Physics and Technology, and Key Laboratory of HEDP of\\ the Ministry of Education, CAPT, School of Physics, Peking University, Beijing 100871, China}
\affiliation{Beijing Laser Acceleration Innovation Center, Huairou, Beijing, 101400, China}

\date{\today}
\begin{abstract}
We develop a theoretical model for self-generated magnetic islands in relativistic laser-driven channels in near-critical-density plasmas.
The islands arise from the nonlinear superposition of the quasi-static magnetic fields generated by the longitudinal channel current $j_x$ and the laser-front-driven transverse current $j_y$.
By deriving the critical conditions among laser depletion, transversely symmetric channel formation, and magnetic-island formation, we identify the laser-plasma parameter window in which the magnetic island structures can exist.
%
Within this window, the balance between the laser ponderomotive force and the charge-separation force, expressed through an effective electron density $n_\mathrm{eff}$, determines the transverse island width $H$, whereas the mismatch between the laser group and phase velocities determines the longitudinal period $L$.
Large-scale particle-in-cell simulations over a broad range of laser intensities and plasma densities validate the resulting scaling laws.
The model turns the island geometry from a qualitative feature of the channel field into a predictable quantity, providing a basis for tailoring electron transport, particle acceleration, high-energy radiation, and novel fusion ignition schemes in relativistic laser-plasma interactions.
\end{abstract}

\maketitle
Relativistic laser-plasma interactions are central to plasma-based accelerators, compact radiation sources, inertial-confinement fusion, and laboratory studies of high-energy-density and astrophysical plasmas~\cite{Tajima1979PRL,Tabak1994POP,Pukhov2003RPP,faure2004laser,Mourou2006RMP,Remington2006RMP,Esarey2009RMP,schaeffer2023proton}.
When a relativistically intense laser pulse (\(I>10^{18}\,\mathrm{W/cm^2}\)) irradiates a near-critical-density target (\(0.1<n_e/n_c<10\)), relativistic transparency and self-focusing can sustain propagation deep into the plasma and excavate a long-lived channel with a transverse size comparable to the laser focal spot~\cite{Pukhov1996PRL,Pukhov1999POP,Chen2007POP,Esarey2009RMP}.
The laser-driven axial electron current then generates an intense azimuthal quasi-static magnetic field~\cite{Pukhov1996PRL,Pukhov2003RPP}. Far from being a passive by-product, this field acts as a magnetic guide: it keeps energetic electrons inside the laser field, rotates their transverse momentum, and changes their dephasing from the optical field, thereby regulating direct laser acceleration (DLA)~\cite{Pukhov1999POP,Gahn1999PRL,Liu2013PRL,Gong2019SR,Arefiev2020PRE,Gong2020PRE,Wang2020POP,Rinderknecht2021NJP,Yeh2021NJP,Hussein2021NJP,Babjak2024PRL,Arefiev2024POP,Tang2024NJP,Valenta2024PRE,Cohen2024SciAdv,Tang2025POP,Lan2025CPB,Yeh2025POP,Bhakta2026NJP,Hadjisolomou2026PRE,Cohen2026APL}. The same deflection drives betatron motion and determines where, in what direction, and at what energy x/\(\gamma\)-ray photons are emitted~\cite{Ji2014PRL,Ji2014POP,Stark2016PRL,Huang2016PRE,Gong2018PPCF,Jansen2018PPCF,Wang2020PRA,Gong2022PRR,Hadjisolomou2022SciRep,Yeh2024POP,Meir2024PRA,Yu2024RMPP,Tangtartharakul2025NJP,Meir2026CommPhys,Yi2026POP}. The deflection also redirects fast-electron transport and the current flow that feed subsequent ion-acceleration stages~\cite{Pukhov2003RPP,Nakamura2010PRL,Bulanov2010POP,Bin2018PRL,Park2019POP,Reichwein2021PPCF,Wang2021PRX,Reichwein2022PRAB,Liu2022PRL,Gong2022PRR2,Lezhnin2022PRR,Qiu2026UltrafastSci}, generation of polarized particle beams~\cite{Gong2021PRL,Reichwein2025RPP}, and exploration of laboratory astrophysical magnetic reconnection~\cite{Gu2019SciRep,Zhang2022CPB,Zhang2024PPCF,Yin2025PPCF,Zhang2026PPCF}. 
Resolving the field structure inside the channel is therefore essential for predicting the electron orbits from which these particle and radiation observables emerge.

\begin{figure*}[t!]
\centering
\includegraphics[width=0.9\textwidth,height=0.45\textheight,keepaspectratio]{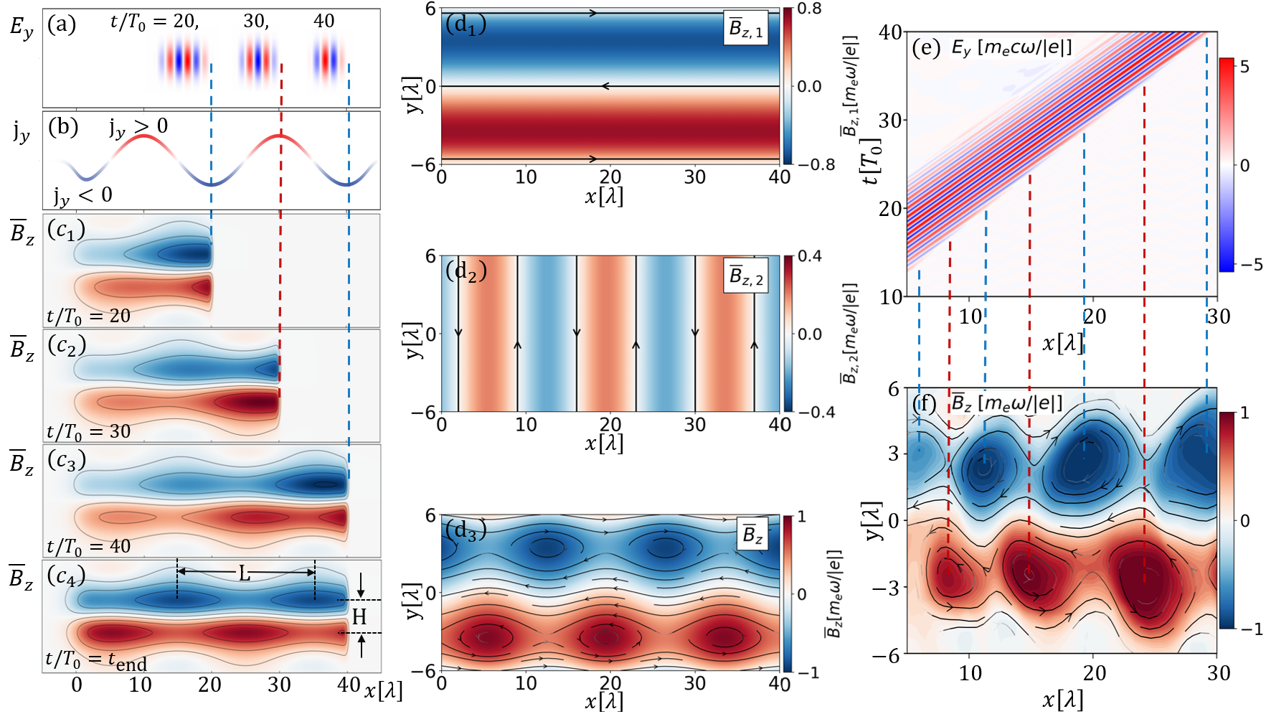}
\caption{Schematic of the magnetic island formation process. ($a$) Laser envelopes in the $x$-$y$ space at $t/T_0=20, 30, 40$. ($b$) Transverse current $j_y$ (at $y=0$) averaged over time versus $x$. ($c_1$)-($c_3$) Island-like structures of the averaged magnetic field $\overline{B}_z$ formed in the $x$-$y$ space over two periods at $t/T_0=20, 30, 40$. ($c_4$) the magnetic field $\overline{B}_z$ averaged over all times after the laser leaves the plasma. ($d_1$) Dominant magnetic field component $\overline{B}_{z,1}$ (background color) and the corresponding longitudinal current $j_x$ (black streamlines). ($d_2$) secondary magnetic field component $\overline{B}_{z,2}$ (background color) and the corresponding transverse current $j_y$ (black streamlines). ($d_3$) their superposition $\overline{B}_{z} = \overline{B}_{z,1} + \overline{B}_{z,2}$ (background color) and the total current $j_{x,y}$ (black streamlines). (e)(f) 2D PIC simulation results for $a_0=30,n_e=0.3n_c$. (e) temporal evolution of the laser electric field $E_y$ at $y=0$. (f) spatial distribution of the quasi-static magnetic field $\overline{B}_z$, where the solid black lines with arrows indicate the current density distribution $j_{x,y}$. The dashed line between (e) and (f) indicates the relation between the spatial distributions of the laser wavefront $E_y$ and the magnetic islands $\overline{B}_z$.}
\label{fig:schematic}
\end{figure*}

However, for a long time theoretical models have mostly simplified the magnetic field inside the channel to a first-order quasi-static magnetic field \(\overline{B}_{z,1}\) with an antisymmetric linear distribution produced solely by the longitudinal current \(j_x\)~\cite{Pukhov1999POP,Pukhov2003RPP}. Such a simplified physical picture cannot perfectly reflect the physical structure of the magnetic field inside the channel.
Earlier studies showed complex magnetic structures such as electron vortices and electromagnetic solitons produced in the interaction of s-polarized laser pulses with plasmas. These structures originate from the quasi-static magnetic wake left in the plasma after the relativistic self-focusing of the laser pulse, where the electron vortices can be described by the Hasegawa-Mima equation, and their antisymmetric vortex arrays are analogous to the K\'arm\'an vortex street in fluid mechanics, often accompanied by nonlinear structures such as post-solitons~\cite{Bulanov1996PRL,Chen2007POP,Yue2022CPB}. These works, however, focused on fluid instabilities or solitary-wave analyses, such as the stability of vortex arrays and the bending instability of laser self-guided channels~\cite{bulanov1997stability,Chen2007POP}, and did not provide a systematic physical explanation of the microscopic magnetic field structures induced by the carrier etching effect of p-polarized lasers.
Only recently, Gong et al., while studying the electron spin polarization in the interaction of ultrarelativistic lasers with critical-density plasmas, clearly explained a nontrivial new structure inside the channel---the ``magnetic islands'' driven by p-polarized laser pulses~\cite{Gong2021PRL}. These magnetic islands originate neither from the tearing-mode instability in magnetohydrodynamics nor from vortex waves. Instead, they are formed by the coupling between the longitudinal current \(j_x\) and the transient transverse current \(j_y\) etched by the laser wavefront at the channel edge. 
The longitudinal current produces the dominant antisymmetric background field \(\overline{B}_{z,1}\), while the transverse current induces a second-order magnetic field \(\overline{B}_{z,2}\) that is locked to the carrier-envelope phase (CEP) of the laser pulse and oscillates periodically along the propagation direction. The nonlinear superposition of the two components forms the spatially interleaved, vertically asymmetric island-like magnetic field distribution shown in Fig.~\ref{fig:schematic}, endowing the channel magnetic field with a completely new fine spatiotemporal structure~\cite{Gong2021PRL}. 
Latest studies further show that such asymmetric magnetic islands can influence the energy transport and particle dynamics in plasma channels. Regarding $\gamma$-ray radiation control, high-energy photons tend to be generated at the edges of the magnetic islands, which is modulated by the CEP~\cite{Cai2025POP}. Regarding particle acceleration, the asymmetry of the magnetic islands can deflect the electron motion asymmetrically, which in turn drives, behind the target, an asymmetric collisionless electrostatic field whose direction can be reversed by a CEP \(\pi\) shift, ultimately enabling active control of the emission direction of high-energy proton beams~\cite{Qiu2026UltrafastSci}.

Although the above works have made certain progress in the discovery of magnetic islands and their physical effects, two key questions remain to be answered before moving from basic understanding to experimental design. First, do the magnetic island structures identified through electron spin polarization under ultrarelativistic laser conditions~\cite{Gong2021PRL} universally exist in the more general near-critical-density plasma channels driven by relativistic intense laser pulses? Second, a theoretical relation between the characteristic geometric sizes of the magnetic islands (such as the transverse width \(H\) and the longitudinal period length \(L\)) and the initial laser-plasma parameters $(a_0, n_e)$ has not yet been constructed from fundamental physical principles. 
The absence of such scaling laws means that we cannot actively design the magnetic island structures to meet specific needs.
For example, in DLA and high-flux x/\(\gamma\) radiation sources, the size of the magnetic islands directly determines their ability to confine and deflect electrons, potentially affecting the quality of the accelerated electron beams as well as the directionality and brightness of the radiation sources~\cite{Huang2016PRE,Arefiev2020PRE,Wang2020PRA,Gong2022PRR,Hadjisolomou2022SciRep,Yeh2024POP,Arefiev2024POP,Meir2024PRA,Cai2025POP,Bhakta2026NJP,Hadjisolomou2026PRE,Meir2026CommPhys}. In laboratory astrophysics, asymmetric magnetic islands that can be precisely characterized provide an ideal platform to simulate asymmetric magnetic reconnection, and the spatial geometry of the islands potentially affects the length of the reconnection current sheet and the reconnection rate~\cite{Gu2019SciRep,Zhang2022CPB,Zhang2024PPCF,Yin2025PPCF,Zhang2026PPCF}. In ion acceleration, the transverse extent and longitudinal periodicity of the islands potentially determine the ability of the induced asymmetric shock structures to control the deflection direction and efficiency of proton beams~\cite{Silva2004PRL,Bin2018PRL,Wang2021PRX,Gong2022PRR2,Qiu2026UltrafastSci}. 
Therefore, establishing a quantitative theoretical model of the geometric sizes of the magnetic islands is the key to unlocking their application potential in the above fields.

Here we address the universality and geometric scaling of these magnetic islands using large-scale 2D particle-in-cell (PIC) simulations and analytical modeling. We first delimit the parameter regions corresponding to laser depletion, a transversely symmetric channel, resolvable magnetic islands, and islands whose period exceeds the target length. 
We then derive \(H\) from the balance between the laser ponderomotive force and the charge-separation force governed by an effective density $n_\mathrm{eff}$, and derive \(L\) from the mismatch between the laser group and phase velocities. The resulting scalings with the similarity parameter \(a_0/n_e\)~\cite{Gordienko2005POP,Pukhov2006PTRSA,Huang2016PRE} agree with the simulations and provide a quantitative basis for designing the internal magnetic structure of relativistic plasma channels.

We use the EPOCH code~\cite{arber2015contemporary} to simulate the interaction of p-polarized laser pulses with near-critical-density plasma targets in two dimensions. 
In the reference case, the laser pulse has a wavelength of $\lambda=1\,\mu m$, a focal spot radius of $w_0=3\,\mu m$, a duration of $\tau_{d}=15T_0\sim50$fs with laser period $T_0=\lambda/c$ and a $\sin ^{2}(\pi t / \tau_{d})$ temporal envelope, and a peak intensity of $1.23 \times 10^{21}\mathrm{W/cm}^2$ (corresponding to a normalized intensity of $a_0=30$). The transverse electric field of the laser pulse is along the $y$-direction.
The target thickness is $l=40\,\mu m$ and the electron density is $n_e=0.3n_c$ (carbon ion density $n_i=n_e/6$), where $n_{c}= \varepsilon_0m_e \omega^2 /e^2$ is the critical plasma density at the laser frequency $\omega=ck$. $m_e$ and $e$ are the electron mass and charge, respectively, and $c$ is the speed of light. The simulation box measures 50$\lambda$ $\times$ 30$\lambda$ (in the $X \times Y$ directions) and is divided into $2000 \times 1000$ cells, with 40 electrons and 20 carbon ion macroparticles per cell. In the simulations, we neglect the generation of radiation photons during the electron motion, but include the reaction force of radiation damping on the electron motion. 
In the following, an averaged quantity is defined as the average of that quantity from the beginning to the end of the simulation, e.g., $\overline{B}_z=\int_{0}^{t_{\mathrm{end}}}B_{z}d\tau/t_{\mathrm{end}}$. In particular, in Fig.~\ref{fig:schematic}($c_1$)-($c_3$), $\overline{B}_z$ denotes the average over the current $2T_0$ interval. 
The normalized parameters are defined as $\hat{x}=x/\lambda,v=v/c,p=p/(m_ec),\hat{n}_e=n_e/n_c.$

\begin{figure*}
\centering
\includegraphics[width=0.98\textwidth]{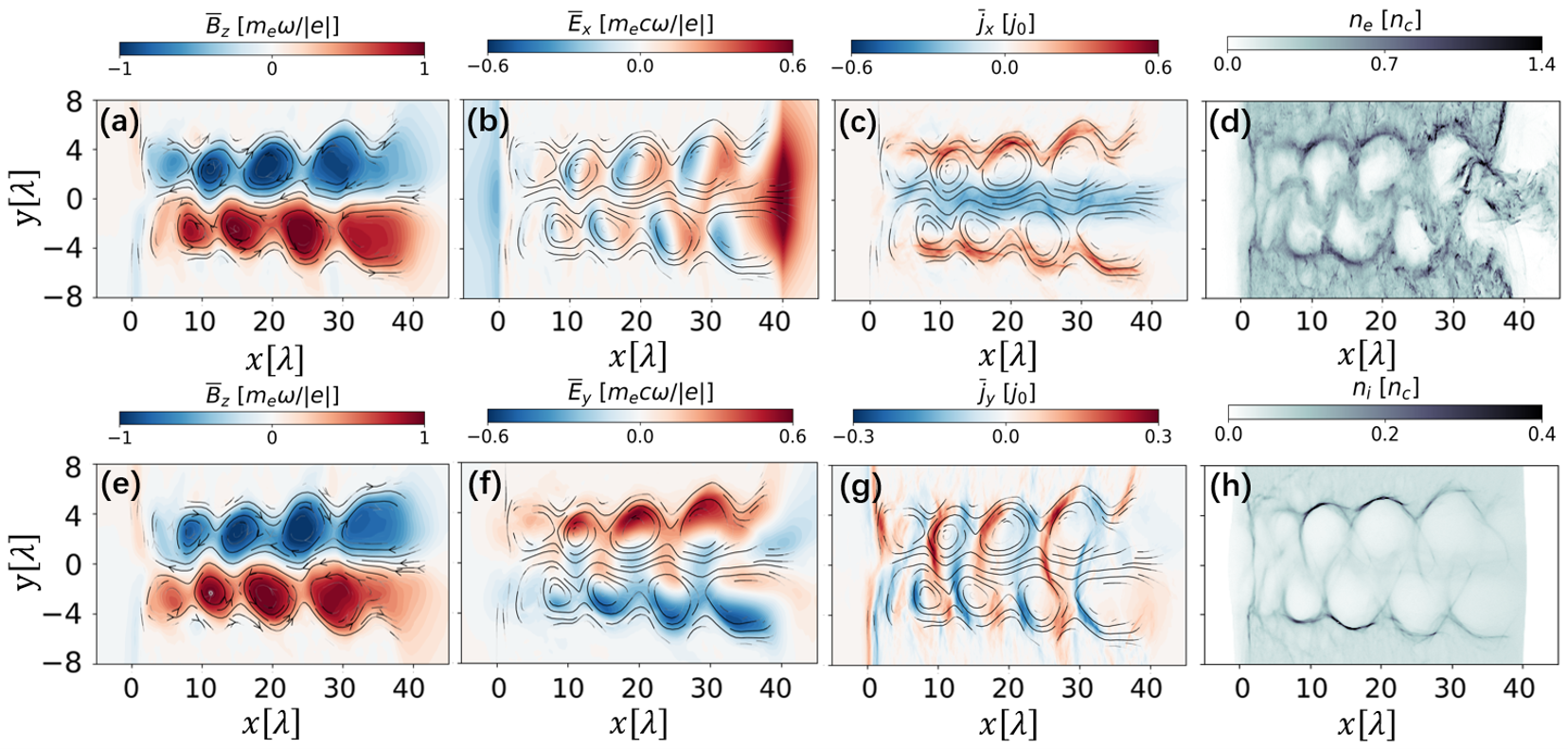}
\caption{PIC simulation results, $a_0=30,\ n_e=0.3n_c$. The spatial distribution of (a) averaged transverse magnetic field $\overline{B}_z$ and current $j_{x,y}$ with $\phi_{0}=0$ and (e) averaged transverse magnetic field $\overline{B}_z$ and current $j_{x,y}$ with $\phi_{0}=\pi$, (b) averaged longitudinal electric field $\overline{E}_x$ and (f) averaged transverse electric field $\overline{E}_y$, (c) averaged longitudinal currents $\overline{j}_x$ and (g) averaged currents $\overline{j}_y$, (d) electron density $n_e$ and (h) ion(Carbon) density $n_i$.}
\label{fig:mi_eybzjne}
\end{figure*}

Figure~\ref{fig:schematic} illustrates the physical process of magnetic island formation.
When a relativistically intense laser pulse (with a normalized amplitude \(a_0 \gg 1\)) interacts with a near-critical-density (\(n_e \sim n_c\)) plasma target, the transverse ponderomotive force of the laser expels the electrons radially, while the ions remain nearly stationary owing to their large mass, thereby forming a strong charge-separation field and a low-density plasma channel inside the target ~\cite{Pukhov1996PRL,Pukhov2003RPP}. In the classical channel model, as the laser pulse propagates, the collective longitudinal motion of the electrons inside the channel dominates the generation of the quasi-static self-generated magnetic field (QPMF)~\cite{Pukhov1999POP,Pukhov2003RPP,Qiao2005POP,Russell2025PRR,Chintalwad2025PRE}.

For p-polarized drivers, however, the QPMF is not simply a linear antisymmetric background. Instead, it exhibits an asymmetric island-like structure---the ``magnetic islands''~\cite{Gong2021PRL}. The concrete formation process is shown in Fig.~\ref{fig:schematic}: when the laser pulse enters and begins to etch the target [Fig.~\ref{fig:schematic}(a)], a dense electron layer accumulates in the front region of the plasma channel, accompanied by a strong transient transverse current \(j_y\) [Fig.~\ref{fig:schematic}(b)]. Owing to the difference between the phase velocity $v_{p}$ and the group velocity $v_g$ of the laser pulse propagating in the plasma, the laser wavefront is depleted while etching the plasma~\cite{Nerush2009PRL}, so that the wavefront exhibits a periodic positive-negative current variation similar to a sine function.
The periodically oscillating electric field $E_y$ of the laser wavefront acts on the electrons accumulated at the channel front $x \sim v_g t$, imparting a transverse momentum $p_y \sim A_y =a_0 \cos(\xi + \phi_0)$ to them, where $\xi = \omega t - k x$ and $\phi_0$ is the carrier-envelope phase. This generates the transient transverse current $j_y \approx-\left|e\right|\int{n_\mathrm{eff}\delta(x/v_g-t)v_ydt}\approx\left|e\right|n_\mathrm{eff} \cos[\omega(x/v_g-x/v_{p}) + \phi_0]\approx\left|e\right|n_\mathrm{eff} \cos(k_2 x + \phi_0)$, where $k_2 = k(v_{p}-v_g)/v_g$ is the effective wave number introduced by the mismatch between the phase and group velocities, and the transverse electron velocity is $v_y=p_y/(\gamma m_e c)\sim \cos(\xi+\phi_0)$. The physical meaning of $n_\mathrm{eff}\delta(x/v_g-t)$ is that the transient transverse current is produced by the transverse motion of the effective-density electron accumulation layer located at the laser front and the channel edge $x \sim v_g t$.
At the wavefront, the electric field is $E_y = -\partial A_y/\partial t \propto \sin(k_2x+\phi_0)$. The comparison shows that $j_y$ leads $E_y$ by $\pi/2$, consistent with the correspondence between the laser envelope $E_y$ and the current $j_y$ shown in Figs.~\ref{fig:schematic}(a)(b). 
On this basis, from Amp\`ere's law $\partial \overline{B}_{z,2}/\partial x = -\mu_0 j_y$, the secondary magnetic field is $\overline{B}_{z,2} \propto -\sin(k_2 x + \phi_0)$, i.e., $\overline{B}_{z,2}$ is spatially in antiphase with $E_y$ at the wavefront. As the ponderomotive force continuously expels the bulk electrons and a quasi-static force balance is reached, a stable plasma channel forms [Figs.~\ref{fig:schematic}($c_1$)-($c_3$)]. During this process, the periodic transverse current \(j_y\) acts together with the collective longitudinal current \(j_x\) inside the channel, coupling to form the magnetic island structures [Fig.~\ref{fig:schematic}($c_4$)]. The magnetic islands etched by the periodic asymmetric transverse current $j_y$ therefore naturally possess an asymmetric periodic structure.

Equivalently, the magnetic islands can be decomposed into \(\overline{B}_{z} = \overline{B}_{z,1} + \overline{B}_{z,2}\). 
The dominant component \(\overline{B}_{z,1}\) originates from the axial current \(j_x\) and has the approximately linear, antisymmetric transverse profile shown in Fig.~\ref{fig:schematic}($d_1$).
The secondary component \(\overline{B}_{z,2}\), shown in Fig.~\ref{fig:schematic}($d_2$), is generated by the transient transverse current \(j_y\) driven by the laser wavefront. 
Owing to the mismatch between the group velocity \(v_g\) and the phase velocity \(v_p\) of the laser in the plasma, the oscillating electric field at the laser wavefront couples to $j_y$ through the electron momentum and directly etches the envelope phase onto the transverse current [Figs.~\ref{fig:schematic}(a)(b)], so that \(\overline{B}_{z,2}\) exhibits, along the propagation direction \(x\), a periodic oscillation with wave number $k_2$ strictly locked to the laser CEP. 
The superposition of the two components breaks the symmetry of the channel magnetic field, ultimately presenting in space a nonlinear, island-like magnetic field structure $\overline{B}_{z}$ outlined by the electron density distribution [Fig.~\ref{fig:schematic}($d_3$)], whose essence is directly related to the kinks and vortex structures of the currents in the plasma channel.
Figures~\ref{fig:schematic}(e) and (f) further verify this phase-locking mechanism directly: each oscillation period of the spatiotemporal envelope of the laser wavefront ($E_y$ at $y=0$) corresponds exactly to one magnetic island, and the two are periodically synchronized in antiphase along the $x$ direction---it is Amp\`ere's law that converts the $\pi/2$ lead of $j_y$ over $E_y$ into the fixed spatial antiphase relation between $\overline{B}_{z,2}$ and $E_y$, thereby etching periodically arranged asymmetric magnetic islands at the channel front.

To verify the above physical picture, Fig.~\ref{fig:mi_eybzjne} presents the full 2D PIC simulation results. The simulations show the dense electron layer accumulated at the channel front [Fig.~\ref{fig:mi_eybzjne}(d)] and the carbon ion distribution acting as the quasi-static background [Fig.~\ref{fig:mi_eybzjne}(h)]. This charge separation directly produces the longitudinal electric field $\overline{E}_x$ inside the channel [Fig.~\ref{fig:mi_eybzjne}(b)]. Meanwhile, the transverse electric field $\overline{E}_y$ observed in the simulation [Fig.~\ref{fig:mi_eybzjne}(f)] accurately captures the spatiotemporal evolution of the laser wavefront. It is this periodic wavefront that modulates the accumulated electrons and generates the strong transient transverse current $\overline{j}_y$ [Fig.~\ref{fig:mi_eybzjne}(g)]. Inside the channel, the longitudinal current $\overline{j}_x$ [Fig.~\ref{fig:mi_eybzjne}(c)] dominates the overall magnetic field background and couples with the periodic $\overline{j}_y$, self-consistently evolving into the asymmetric magnetic island structures [Fig.~\ref{fig:mi_eybzjne}(a)]. When the laser CEP  is changed from $\phi_0=0$ to $\pi$, the spatial distribution of the magnetic islands flips symmetrically along the $x$ direction [Fig.~\ref{fig:mi_eybzjne}(e)], further confirming the intrinsic connection between the magnetic islands and the laser wavefront etching effect. 
Since the magnetic island structures are asymmetrically distributed in the transverse direction, electrons in different transverse regions experience different instantaneous magnetic field strengths, which leads to spatial inhomogeneity of relevant consequence, such as radiative spin flips inside the channel~\cite{Gong2021PRL}. Therefore, establishing a quantitative scaling law of the island size is also a physical prerequisite for further actively designing and controlling strong-field QED effects~\cite{DiPiazza2012RMP,Jansen2018PPCF,Samsonov2022MRE,Wan2023MRE,Blackburn2023POP,Cai2025POP,Reichwein2025RPP,Liu2026MRE}.

\begin{figure*}[t!]
\centering
\includegraphics[width=0.7\textwidth]{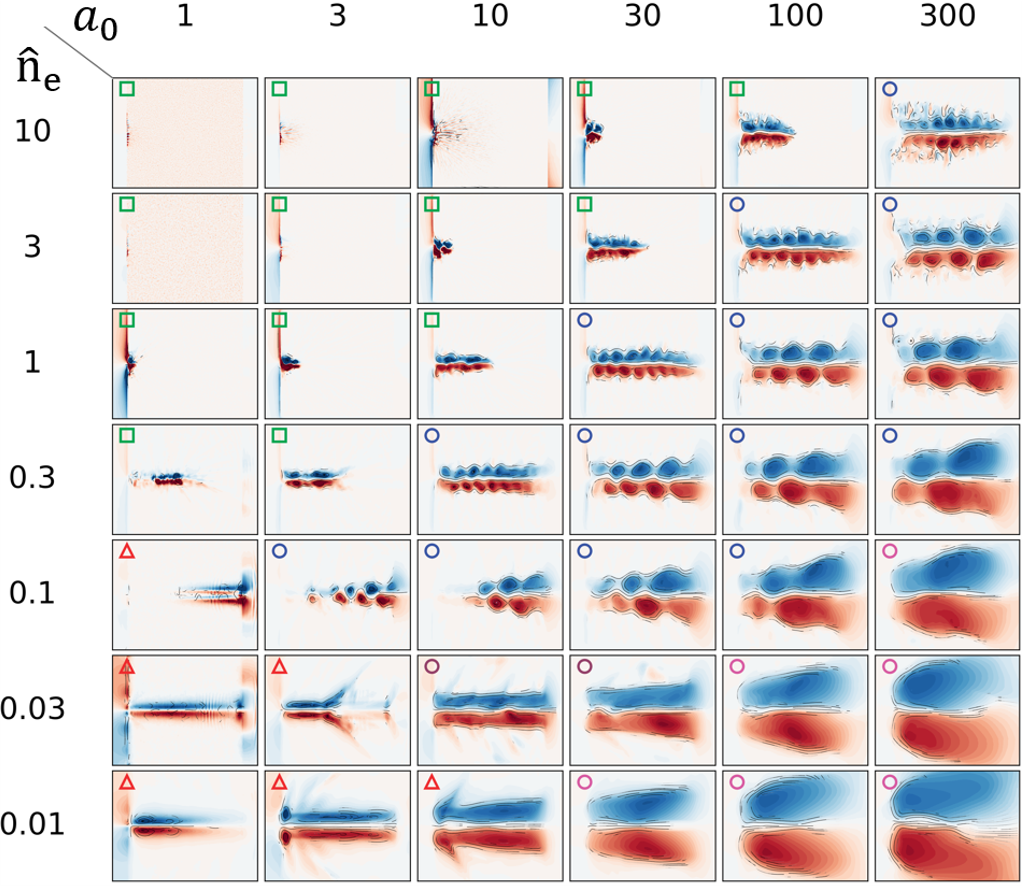}
\caption{Time-averaged magnetic field $\overline{B}_z$ for different laser intensities $a_0(1-300)$ and plasma densities $n_e(0.01-10)$; the upper limit of the color bar in each panel is $|B_z|_\mathrm{max}$. Green squares: the laser cannot penetrate the target, and no magnetic islands form; red triangles: the laser penetrates the target but forms only a transversely symmetric channel, without magnetic islands; blue circles: magnetic islands form; pink circles: magnetic islands form but a single island length exceeds the target length; mulberry circles: the transition between the transversely symmetric channel and the magnetic-island channel}
\label{fig:bz_multi} 
\end{figure*}

\begin{figure*}[t!]
\centering
\includegraphics[width=0.8\textwidth]{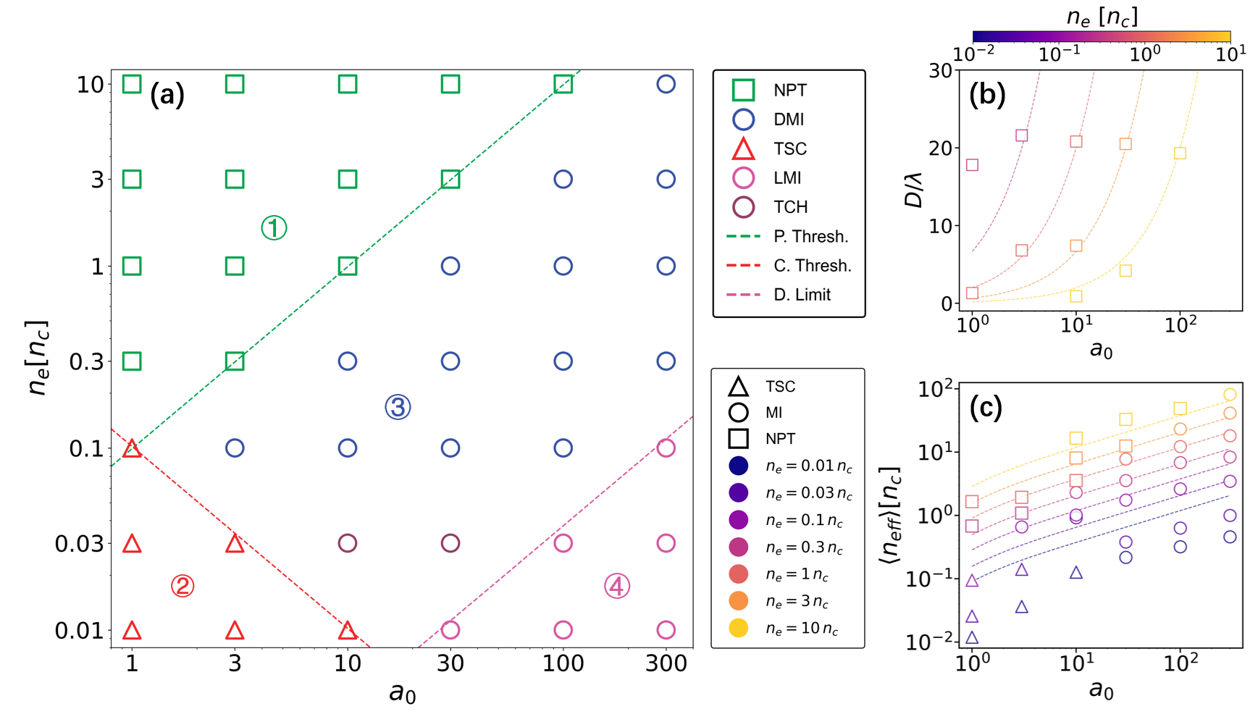}
\caption{(a) Magnetic island formation for different laser intensities $a_0(1-300)$ and plasma electron densities $\hat{n}_e(0.01-10)$. Green squares: the laser pulse is fully depleted without penetrating the target (NPT); red triangles: the laser pulse penetrates the target but forms only a transversely symmetric channel (TSC); blue circles: resolvable magnetic islands form (DMI); pink circles: magnetic islands form but a single island length exceeds the target length (LMI); mulberry circles: the transition between the transversely symmetric channel and the magnetic-island channel (TCH). The green dashed line is the boundary between the NPT and DMI regions, the red dashed line is the boundary between the TSC and DMI regions, and the pink dashed line is the boundary between the DMI and LMI regions. (b) Penetration depth $D$ versus $a_0$ for different $\hat{n}_e$; (c) effective density $\hat{n}_\mathrm{eff}$ versus $a_0$ for different $\hat{n}_e$. Symbols: PIC simulation data; dashed lines: analytical estimations.}
\label{fig:3}
\end{figure*}

To study the conditions and parameter range for the generation of the magnetic islands, 
we scan the laser intensity $a_0$ from 1 to 300 and, for each value of $a_0$, the plasma density $\hat{n}_e$ from 0.01 to 10, and the results are shown in Fig.~\ref{fig:bz_multi}.
According to the magnetic field distribution, the magnetic structures under different parameters can be classified into four categories: \ding{172}$\,$ when $a_0$ is small and $n_e$ is large, the laser energy is depleted inside the plasma and the magnetic field distribution does not penetrate the target (green squares, Non-Penetrated Target, NPT); \ding{173}$\,$ when $a_0$ is small and $n_e$ is also small, the laser penetrates the target, but instead of forming a transversely asymmetric magnetic-island channel, it forms a transversely symmetric channel (red triangles, Transversely Symmetric Channel, TSC)~\cite{Tajima1979PRL}; the laser penetrates the target and forms a transversely asymmetric channel, in which periodic magnetic islands are visible (circles). Among these cases we can further distinguish: \ding{174}$\,$ magnetic islands whose number and longitudinal period length can be clearly identified (blue circles, Discrete Magnetic Island, DMI); \ding{175}$\,$ the channel magnetic field is close to vertically symmetric, without an obviously vertically asymmetric magnetic island distribution (pink circles, Long Magnetic Island, LMI). 
This case belongs, like the TSC region, to the transversely symmetric channels, but the physical mechanisms differ: in the TSC region, the tail electrons are pulled back by the laser field and re-close, whereas in the LMI region the tail electrons are not pulled back and a transversely asymmetric magnetic-island channel would have formed---it is only because the longitudinal period of the islands exceeds the target length that complete islands cannot be resolved within the target. 
The transition state between the transversely symmetric channel and the magnetic-island channel (mulberry circles, Transition Channel, TCH).
Simplifying the above parameter scan, we obtain our parameter classification diagram shown in Fig.~\ref{fig:3}(a). Clearly, there is a boundary between \ding{172}$\,$NPT and \ding{174}$\,$DMI, which separates two different physical mechanisms.
From the simulation results [green squares in Fig.~\ref{fig:bz_multi}], it can be seen that when the laser intensity is weak and the plasma density is high, i.e., when $\hat{n}_e/a_0$ is large, the laser cannot penetrate the target and naturally no magnetic islands form; whereas relativistic self-transparency occurs during the interaction of relativistic lasers with plasmas, and the laser can penetrate plasma targets with $\hat{n}_e\sim\gamma$; however, we find that when $\hat{n}_e/a_0>0.1$, the laser cannot penetrate the target. Therefore, below we discuss the relation between the penetration depth of the laser in the plasma and the laser-plasma parameters.

Owing to the plasma dispersion, the laser group velocity $v_g$ is smaller than its phase velocity $v_p$. As the pulse propagates forward, the carrier phase continuously slips backward relative to the pulse envelope, so that the strong field at the wavefront is continuously eroded. 
Within the pulse duration, the maximum propagation distance over which the wavefront can be sustained by the laser energy, namely the penetration depth $D$, is determined by this slippage process.
Consider a laser pulse of duration $\tau_{d}$ and spatial length $l_{\text{pulse}} \approx c\tau_{d}$. In the reference frame comoving with the pulse at the group velocity $v_g$, the rate at which the carrier phase at the wavefront slips relative to the pulse envelope is $v_p - v_g$. Therefore, the time required for the wavefront to slip by one full pulse length is $\tau_{s} \approx \frac{c\tau_{d}}{v_p - v_g}$.
Within this time $\tau_{s}$, the distance over which the pulse as a whole propagates forward at the group velocity $v_g$
is the maximum penetration depth: $D \sim v_g \tau_{s}\approx [c\tau_{d}/(v_p - v_g)]\cdot{v_g}$. In the plasma, the group velocity of the laser is $v_g=\frac{d\omega}{dk}\approx c\sqrt{1-\frac{\omega_p^2}{\omega^2}}$ and the phase velocity is $v_p=\frac{\omega}{k}\approx c/\sqrt{1-\frac{\omega_p^2}{\omega^2}}$. Due to the relativistic effect, the plasma frequency is $\omega_p=\sqrt{\frac{n_ee^2}{\gamma\varepsilon_0m_e}}=\sqrt{\frac{\hat{n}_e\omega^2}{\gamma}}$. Therefore, $v_g\approx c\sqrt{1-\frac{n_e}{\gamma}}\approx c(1-\frac{n_e}{2\gamma})$, and the phase velocity is $v_p\approx c/\sqrt{1-\frac{n_e}{\gamma}}\approx c(1+\frac{n_e}{2\gamma})$. For a linearly polarized laser, $\gamma\sim\sqrt{1+a_0^2/2}$.
We thus obtain the scaling relation of the penetration depth:
\begin{equation}
D \approx \frac{\gamma}{\hat{n}_e}\,c\tau_{d}\sim\frac{a_0}{\sqrt{2}\hat{n}_e}\,c\tau_{d}.
\label{eq:D_scaling}
\end{equation}

To penetrate a target of thickness $l$ requires $D \gtrsim l$. With $D$ given by Eq.~\eqref{eq:D_scaling}, entering the \ding{174}DMI region from the \ding{172}NPT region is determined by the condition
\begin{equation}
\frac{a_0}{\hat{n}_e}\gtrsim\frac{\sqrt{2}\,l}{c\tau_d},
\label{eq:boundary_NPT_DMI}
\end{equation}
which, for our simulation parameters, corresponds to $a_0/\hat{n}_e\gtrsim4$. Since during the actual interaction the laser is not always at its peak intensity $a_0$ and the average intensity is approx $\frac{1}{\tau_{d}}\int_{0}^{\tau_{d}}a_0{\sin^2(\pi t/\tau_{d})}dt=a_0/2$, in the competition between the \ding{172}NPT and \ding{174}DMI regions, $a_0/\hat{n}_e\gtrsim10$ corresponds to entering the \ding{174}DMI region and $a_0/\hat{n}_e\lesssim10$ to entering the \ding{172}NPT region, consistent with the simulation results [green boundary in Fig.~\ref{fig:3}(a)].
We further compare the deposition depth inside the target for the NPT data with the theoretically calculated $D$, and they agree reasonably well [Fig.~\ref{fig:3}(b)].

Next we consider the boundary between \ding{173}TSC and \ding{174}DMI.
The difference between the two lies in whether the formed channel remains transversely symmetric. 
Whether these transient asymmetric structures can develop into quasi-static magnetic islands is decided by whether the expelled electrons can be pulled back into the evacuated region within half a laser period, before they pile up at the channel edge. 
The pull-back is driven by the symmetric ponderomotive force of the laser: if the electrons are pulled back in time, no persistent pile-up forms at the channel edge, the asymmetric current $j_y$ remains transient and leaves no quasi-static signature, and the evacuated region re-closes at the tail of the laser, remaining transversely symmetric and forming a transversely symmetric channel propagating together with the laser~\cite{Nerush2009PRL}. 
If, instead, the electrons cannot be pulled back, i.e., once the asymmetric force overcomes the symmetric ponderomotive force, they pile up persistently at the channel edge with effective density $n_\mathrm{eff}$, sustaining the asymmetric current $j_y$ and the $\overline{B}_{z,2}$. The charge-separation force produced by the expelled electrons and the transverse ponderomotive force of the laser then reach a quasi-static force balance, and the electrons do not re-inject at the laser tail. 
As the ion response time (ps) is much longer than the laser pulse duration, the ion background is exposed; the electrons cannot immediately flow back to refill after the laser passes, but instead release back slowly, leaving behind a transversely asymmetric quasi-static plasma density channel, on which $\overline{B}_{z,2}$ superimposes on $\overline{B}_{z,1}$ to form the magnetic islands. 
In this criterion, we consider the dynamics of the electrons near the tail of the evacuated region (behind the wavefront that etches the plasma). 
At this position the wavefront has already passed, so there is no dense electron accumulation layer driven by the laser front, and the quasi-static restoring field experienced by the expelled electrons is mainly provided by the exposed background ions. Below we use a simple model to delineate these two physical mechanism regimes.

We assume that after the laser pulse enters the plasma target and expels the electrons, an evacuated channel is locally formed regardless of whether the electrons subsequently re-inject and whether a transversely symmetric or asymmetric channel ultimately forms. When examining the criterion for the electron backflow at the tail of the evacuated channel, since the ions barely move owing to their large mass, the tail region of the evacuated channel mainly exposes the background ions, forming a positive charge background composed of ions. Therefore, the net charge density $\hat{n}(0)$ inside the evacuated channel at this moment is taken as the unperturbed background ion density, i.e., $\hat{n}(0)\approx Z\hat{n}_i = \hat{n}_e$. The transverse width $H$ of the evacuated channel is determined by the balance between the transverse pressure of the laser and the electrostatic pressure arising from the charge separation of these background ions~\cite{faure2004laser}.
The laser pulse propagates along the $x$ direction and forms a one-dimensional slab channel in the transverse ($y$) direction. The net charge density of the plasma at $y=0$ is $n(0)$, and the ions are immobile. The laser is linearly polarized with a normalized vector potential envelope $a$, and the relativistic factor of the electrons in the laser field is $\gamma \sim \sqrt{1 + a^2/2}$.
Assuming that the plasma reaches a quasi-static state and neglecting the thermal pressure, the electron momentum equation is $m_e c^2 \nabla \gamma \approx e \nabla \Phi,$
where $\Phi$ is the electrostatic potential produced by the charge separation. Combined with Poisson's equation $\nabla^2 \Phi =  e (n_e - n(0)) /\varepsilon_0,$
this yields the relation between the electron density and $\gamma$, $\hat{n}_e \approx \hat{n}(0) + \nabla^2 \gamma$.
When $a_0$ is sufficiently large, the electrons inside the channel are completely evacuated, i.e., $n_e \approx 0$, and the relation above reduces to the constraint $\nabla^2 \gamma \approx -\hat{n}(0)$ inside the evacuated region.

Let the full width of the channel be $H$, with the electrons completely evacuated in $-H/2 < y < H/2$. For the one-dimensional slab, $\nabla^2 \gamma \approx \dfrac{d^2\gamma}{dy^2}$. Integrating the constraint $\nabla^2\gamma\approx-\hat{n}(0)$ across the channel and using the symmetry $\left.\dfrac{d\gamma}{dy}\right|_{y=0}\approx0$ gives $\gamma(y) \approx \gamma(0) - \dfrac{\hat{n}(0)}{2} y^2$.
At the channel boundary $y \approx \pm H/2$, the laser field decays to zero and the electrons return to the unperturbed state $\gamma \sim 1$. Therefore $
\gamma\left(\frac{H}{2}\right) \approx \gamma(0) - \frac{\hat{n}(0)}{2}\left(\frac{H}{2}\right)^2 \approx 1.$
At the channel center, $\gamma(0) \sim \sqrt{1 + a_0^2/2}$; substituting this gives $\sqrt{1 + a_0^2/2} - 1 \approx {n(0) H^2}/{8},$ i.e.,
\begin{equation}
\hat{H} \sim\kappa_1\sqrt{\frac{8(\sqrt{1+a_0^2/2}-1)}{\hat{n}(0)}},
\label{eq:H_final}
\end{equation}
where $\kappa_1\sim0.7$ is a coefficient accounting for the fact that during the actual interaction the laser is at its peak intensity $a_0$ only part of the time owing to the $\sin^2$ temporal envelope (the actual average intensity is $a_0/2$, $\sqrt{a_0/2}\approx0.7\sqrt{a_0}$). We have thus obtained the relation between the transverse width $H$ of the electron-evacuated region against the ion background and the parameters $a_0,n_e$ after the laser expels the electrons. When the electrons expelled from the axis to the edges at $\pm H/2$ can be pulled back by the laser in time, we consider the evacuated channel as remaining transversely symmetric. If they cannot be pulled back in time, the evacuated channel develops into a transversely asymmetric magnetic-island channel.
Since $\frac{d\xi}{dt}=\omega(\gamma-p_x)/\gamma=\omega R/\gamma,R\equiv\gamma-p_x\sim1$, we have $\frac{dy}{d\xi}=\frac{dy}{dt}\frac{dt}{d\xi}\approx v_y\gamma/\omega=p_y$. Therefore the maximum distance an electron can move within half a period is $y_{\mathrm{max}}\approx\int_{-\pi/2}^{\pi/2}a_0\cos\xi d\xi=2a_0$.
Using $H/2\gtrsim y_{\mathrm{max}}$ as the critical condition separating \ding{173}TSC from \ding{174}DMI, and substituting $y_{\mathrm{max}}=2a_0$ together with $\hat{H}$ from Eq.~\eqref{eq:H_final} with $\hat{n}(0)=\hat{n}_e$, entering the \ding{174}DMI region requires
\begin{equation}
\hat{n}_e\gtrsim\frac{\kappa_1^2\left(\sqrt{1+a_0^2/2}-1\right)}{2a_0^2},
\label{eq:boundary_TSC_DMI}
\end{equation}
which, for our simulation parameters, corresponds to $a_0\hat{n}_e\gtrsim0.1$; conversely, $a_0\hat{n}_e\lesssim0.1$ enters the \ding{173}TSC region. This is the red boundary shown in Fig.~\ref{fig:3}(a).
It should be emphasized that the above derivation, based on the background ion density $\hat{n}_e$, applies only to the initial backflow criterion at the tail of the evacuated region. The actual transverse width $H$ is set by the wavefront-etching process. As the laser propagates, its ponderomotive force pushes electrons toward the channel edge, forming a dense accumulation layer ($n_\mathrm{eff}\gg Z n_i=n_e$) that sustains the asymmetric transverse current $j_y$ and determines the electrostatic pressure $P_{\mathrm{es}}$ balancing the ponderomotive pressure $P_{\mathrm{p}}$. In Eq.~\eqref{eq:H_final}, $\hat{n}(0)$ should therefore be replaced by the effective density $\hat{n}_\mathrm{eff}$ of this accumulation.

As shown in Fig.~\ref{fig:mi_eybzjne}(g) and (d), the wavefront etching drives a density accumulation at the channel front, which we take as the effective density $\hat{n}_\mathrm{eff}$ of the asymmetric transverse current. Balancing the ponderomotive light pressure, $P_{\mathrm{p}}\approx I/c\approx\frac{1}{2}a_0^2m_ec^2n_c$, against the electrostatic pressure of the accumulation, $P_{\mathrm{es}}\approx\frac{1}{2}\varepsilon_0E_{\mathrm{s}}^2\approx\frac{\gamma e^2c^2n_\mathrm{eff}^2}{2\varepsilon_0\omega^2\hat{n}_e}$, where $E_{\mathrm{s}}\approx \frac{en_\mathrm{eff}}{\varepsilon_0}\frac{c}{\omega_p}$ follows from Gauss's law, gives $P_{\mathrm{p}}\approx P_{\mathrm{es}}$, i.e.,
\begin{equation}    
    \hat{n}_\mathrm{eff}\approx\sqrt{\frac{a_0^2\hat{n}_e}{\gamma}}\sim\sqrt{\frac{a_0^2\hat{n}_e}{\sqrt{1+a_0^2/2}}}
    \label{eq:n_eff}
\end{equation}

\begin{figure}[tb]
\centering
\includegraphics[width=0.5\textwidth]{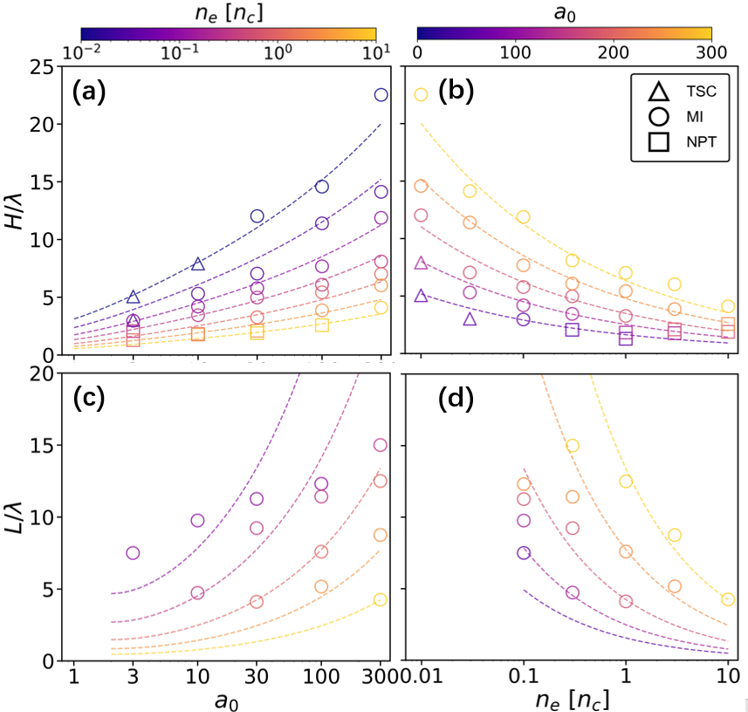}
\caption{Relations of the transverse width $H$ of the channel magnetic field and the longitudinal period length $L$ of the magnetic islands with the laser intensity $a_0$ and the plasma density $n_e$. (a) $H$ versus $a_0$ for different $n_e$; (b) $H$ versus $n_e$ for different $a_0$; (c) $L$ versus $a_0$ for different $n_e$; (d) $L$ versus $n_e$ for different $a_0$. Symbols: simulation data; dashed lines: theoretical fits.}
\label{fig:LH_ver}
\end{figure}

In the simulations, $\hat{n}_{\mathrm{eff}}$ is extracted as the time average of half the maximum electron density at the accumulation, since the charge-separation field there is partly canceled by the background ions and the electrostatic force on a single electron is not always at its maximum.
Comparing Eq.~\eqref{eq:n_eff} with the simulation results [Fig.~\ref{fig:3}(c)], the prediction deviates considerably in the TSC region, where the electrons are pulled back and the pile-up is transient, so that the time-averaged $\hat{n}_{\mathrm{eff}}$ no longer carries the wavefront force-balance meaning of Eq.~\eqref{eq:n_eff}. The use of $\hat{n}_{\mathrm{eff}}$ in Eq.~\eqref{eq:H_final} remains valid, however, $H$ is set by the instantaneous wavefront balance, which holds locally whether or not the pile-up persists. The predicted width therefore still agrees with the TSC-region measurements.
Substituting Eq.~\eqref{eq:n_eff} into Eq.~\eqref{eq:H_final} yields the transverse width of the magnetic islands:
\begin{equation}
    \hat{H} \sim \kappa_1\sqrt{\frac{8(\gamma-1)}{a_0\sqrt{\hat{n}_{e}/\gamma}}}
    \sim\kappa_1\sqrt{\frac{8(\sqrt{1+a_0^2/2}-1)}{a_0\sqrt{\hat{n}_{e}/\sqrt{1+a_0^2/2}}}}.
    \label{eq:end}
\end{equation}
The theoretical results agree well with the simulations [Figs.~\ref{fig:LH_ver}(a)(b)].

In the interaction of lasers with near-critical-density plasmas, the magnetic island structures possess not only transverse scales. Their longitudinal scale also has an important influence on processes such as electron acceleration and x/$\gamma$-ray radiation~\cite{Gong2018PPCF,Jansen2018PPCF,Wang2020PRA,Gong2022PRR,Yeh2024POP,Meir2024PRA,Yi2026POP,Bhakta2026NJP}.
The periodic longitudinal distribution of the magnetic islands originates from the group--phase-velocity mismatch in the channel: each time the pulse advances by one wavelength $\lambda$, the carrier phase at its front slips backward by $(v_p-v_g)\lambda/v_g$ relative to the envelope. When the accumulated slippage reaches $\lambda$, the transverse current $j_y$ completes one spatial oscillation, over a propagation distance $\hat{L}\approx[cT_0/(v_p-v_g)]\,v_g\approx\gamma/\hat{n}_e$; consistently, $\overline{B}_{z,2}\propto-\sin(k_2x+\phi_0)$ yields $L=2\pi/k_2$.
Using the effective pile-up density $\hat{n}_{\mathrm{eff}}$ of Eq.~\eqref{eq:n_eff}, rather than the initial density $\hat{n}_e$, gives the longitudinal length of the magnetic islands:
\begin{equation}    
    \hat{L}\sim \kappa_2\frac{\gamma\sqrt{\gamma}}{a_0\sqrt{\hat{n}_{e}}}\sim\kappa_2\frac{(1+a_0^2/2)^\frac{3}{4}}{a_0\sqrt{\hat{n}_{e}}}
    \label{eq:L}
\end{equation}
where $\kappa_2\sim 1.3$ is a coefficient. When $L$ exceeds the target length $l$, no obvious asymmetric island distribution can be observed. 
Therefore, entering the \ding{175}LMI region is determined by $L\gtrsim l$, which, with $\hat{L}$ from Eq.~\eqref{eq:L} in the limit $a_0\gg1$, reads
\begin{equation}
\frac{a_0}{\hat{n}_e}\gtrsim 2^{3/2}\left(\frac{l}{\kappa_2}\right)^2,
\label{eq:boundary_DMI_LMI}
\end{equation}
which, for our simulation parameters, corresponds to $a_0/\hat{n}_e\gtrsim3\times10^3$; conversely, $a_0/\hat{n}_e\lesssim3\times10^3$ enters the \ding{174}DMI region, i.e., the pink theoretical line in Fig.~\ref{fig:3}(a).
This also explains why, when $a_0/\hat{n}_e$ is large, the simulation results of island length $L$ are smaller than the theoretical prediction [Figs.~\ref{fig:LH_ver}(c)(d)]. In this case the target is not long enough, and the simulated magnetic islands are compressed to within the target length, while the theoretical island length is not long enough to exceed the whole target length.
It is worth emphasizing that $D$ and $L$ originate from the same mechanism, i.e. the group and phase velocity mismatch. $D$ is the propagation distance over which the carrier phase slips by one pulse length $c\tau_d$ relative to the envelope, and $L$ that over which it slips by one laser wavelength $cT_0$. Both scale as $\gamma/\hat{n}_e$, so the green and pink boundaries in Fig.~\ref{fig:3}(a) share the same slope of the parameter dependence. For $D$, however, $a_0/\hat{n}_e$ is small, where the electrons are not completely evacuated, the laser interacts mainly with the unperturbed background plasma, and $P_{\mathrm{p}}\approx P_{\mathrm{es}}$ does not hold, so $n_\mathrm{eff}$ is not involved.
For $a_0\gg1$, $\hat{n}_\mathrm{eff}\sim\sqrt{a_0\hat{n}_e}$, $\hat{L}\sim\sqrt{a_0/\hat{n}_e}$, and $\hat{H}\sim(a_0/\hat{n}_e)^{1/4}$. Once the magnetic-island conditions are satisfied, both the transverse width and the longitudinal length are thus determined by the parameter $a_0/\hat{n}_e$, which is same to the self-similar parameter $S\equiv n_e/(a_0 n_c)$ ~\cite{Gordienko2005POP,Pukhov2006PTRSA,Huang2016PRE}.

In summary, this work systematically studies the scaling laws of the geometric sizes of self-generated magnetic islands in the interaction of a relativistically intense laser pulse with near-critical-density plasmas. Based on the dynamic balance between the laser ponderomotive force and the charge-separation electrostatic force governed by the effective electron density, as well as the group- and phase-velocity mismatch of the laser propagation in plasma, we develop a theoretical model of the transverse width $H$ and the longitudinal period length $L$ of the magnetic islands. Through large-scale PIC simulations, the model is verified over a broad range of laser intensities and plasma densities, and the parameter range in which the magnetic island structures can effectively exist is delimited theoretically for the first time. This model not only fills the gap in the quantitative prediction of the magnetic islands, but also opens a way to actively manipulate the magnetic field structure inside the channel.
Actively designing the spatial structures of the magnetic islands based on these scaling laws is expected to further enable active control of the radiation and spin-poarization processes of high-energy electrons in the plasma channel ~\cite{Gong2021PRL,Blackburn2023POP,Yeh2024POP,Cai2025POP,Meir2026CommPhys}. This work provides a key theoretical tool for understanding the complex magnetic field structure in relativistic laser-driven plasma channels and its applications in particle acceleration~\cite{mangles2004monoenergetic,faure2004laser,Esarey2009RMP,Bin2018PRL,Reichwein2021PPCF,Wang2021PRX,Reichwein2022PRAB,Liu2022PRL,Gong2022PRR2,Lezhnin2022PRR,Qiu2026UltrafastSci}, high-energy x/$\gamma$-ray radiation sources~\cite{DiPiazza2012RMP,Stark2016PRL,Meir2024PRA}, laboratory astrophysics~\cite{Gu2019SciRep,Zhang2022CPB,Yin2025PPCF,Zhang2024PPCF,Zhang2026PPCF}, and relativistic-laser-driven fast-ignition fusion~\cite{Tabak1994POP,Cai2009PRL}.

\section{Acknowledgements} 
This work was supported by the Strategic Priority Research Program of the Chinese Academy of Sciences (Grant No. XDB1550100), the National Key R\&D Program of China (2025YFF0515103), the National Natural Science Foundation of China (Grant No. 12675316, No. 11921006), the National Grand Instrument Project (No. 2019YFF01014400), and the Fundamental and Interdisciplinary Disciplines Breakthrough Plan of the Ministry of Education of China under grant No. JYB2025XDXIM204.
The simulations are supported by the High-Performance Computing Platform of Peking University. The code EPOCH is funded by UK EPSRC Grants No. EP/G054950/1, No. EP/G056803/1, No. EP/G055165/1 and No. EP/M022463/1.
Z.G. acknowledges the National Natural Science Foundation of China (NNSFC) Grants No. 12447101, the CAS Project for Young Scientists in Basic Research (Grant No. YSBR-141), and the HPC Cluster of ITP-CAS for providing computational resources.
The authors would like to thank K. Z. Hatsagortsyan, C. H. Keitel, A. V. Arefiev, S. V. Bulanov, A. Pukhov, and I. Y. Kostyukov for useful discussions.

\bibliography{aa}
\end{document}